%% file: main.tex
\documentclass{article}

\PassOptionsToPackage{numbers, compress}{natbib}
\usepackage[preprint]{neurips_2024}

\usepackage[utf8]{inputenc} 
\usepackage[T1]{fontenc}    
\usepackage{hyperref}       
\usepackage{url}            
\usepackage{booktabs}       
\usepackage{amsfonts}       
\usepackage{nicefrac}       
\usepackage{microtype}      
\usepackage{xcolor}         
\usepackage{graphicx}
\usepackage{multirow} 
\usepackage{amsmath}
\usepackage{cleveref}
\usepackage{xspace}
\usepackage{caption}
\usepackage{subcaption}

\title{Is Tokenization Needed for Masked Particle Modelling?}

\author{
Matthew Leigh \\
University of Geneva \\
\texttt{matthew.leigh@unige.ch} \\
\and 
Samuel Klein \\
University of Geneva \\
\texttt{samuel.klein@unige.ch} \\
\and 
Fran\c{c}ois Charton \\
Meta FAIR \\
\texttt{fcharton@meta.com} \\
\and 
Tobias Golling \\
University of Geneva \\
\texttt{tobias.golling@unige.ch} \\
\and
Lukas Heinrich \\
Technical University of Munich \\
\texttt{lukas.heinrich@cern.ch} \\
\and
Michael Kagan \\
SLAC National Accelerator Laboratory \\
\texttt{makagan@slac.stanford.edu} \\
\and
In{\^e}s Ochoa \\
Laboratory of Instrumentation and Experimental Particle Physics, Lisbon \\
\texttt{ines.ochoa@cern.ch} \\
\and
Margarita Osadchy \\
University of Haifa \\
\texttt{rita@cs.haifa.ac.il} \\
}

\newcommand{\im}[1]{\footnotesize{\color{green} $\uparrow #1$}}

\newcommand{\pt}{\ensuremath{p_\text{T}}}
\newcommand{\xc}{\ensuremath{x^\text{c}}\xspace}
\newcommand{\xid}{\ensuremath{x^\text{id}}\xspace}

\begin{document}

\maketitle

\input{body}

\end{document}

%% file: body.tex
\begin{abstract}
 In this work, we significantly enhance masked particle modeling (MPM), a self-supervised learning scheme for constructing highly expressive representations of unordered sets relevant to developing foundation models for high-energy physics. In MPM, a model is trained to recover the missing elements of a set, a learning objective that requires no labels and can be applied directly to experimental data. We achieve significant performance improvements over previous work on MPM by addressing inefficiencies in the implementation and incorporating a more powerful decoder. We compare several pre-training tasks and introduce new reconstruction methods that utilize conditional generative models without data tokenization or discretization. We show that these new methods outperform the tokenized learning objective from the original MPM on a new test bed for foundation models for jets, which includes using a wide variety of downstream tasks relevant to jet physics, such as classification, secondary vertex finding, and track identification.
\end{abstract}

\section{Introduction}
\label{sec:intro}

The field of high-energy physics (HEP) has increasingly integrated machine learning (ML) methods to tackle diverse challenges, including event reconstruction, anomaly detection, and data generation.
These developments have largely mirrored the trends of the wider ML community.
Model sizes across all fields have grown exponentially, and transformer-based neural networks have become the dominant architecture for many tasks.
However, despite some initial studies \cite{mpm,harris2024resim,kishimoto2023pretraining,supervisedfinetuning,birk2024omnijet,mikuni2024omnilearn,zhao2024large,JetCLR},
HEP has yet to truly adopt foundation models (FMs)~\cite{bommasani2022opportunities}, large pre-trained models that can be fine-tuned on many downstream tasks, which are prevalent in the fields of natural language processing (NLP)~\cite{BERT,gpt,gpt4,bart, NEURIPS2020_1457c0d6} and computer vision (CV)~\cite{dino,ramesh2021zeroshot, AlayracDLMBHLMM22, dinov2, mae, beit}.


An FM is exposed to a large corpus of domain-related data with the goal of learning expressive representations of the subject matter.
This is referred to as \textit{pre-training}, and it is usually self-supervised; the model is given input samples but no associated truth labels.
Once pre-trained, FMs are fine-tuned on specific tasks in a supervised manner.
In NLP, typical pre-training tasks consist of predicting the next token in the input sequence (GPT~\cite{gpt}) or predicting randomly masked tokens (BERT ~\cite{BERT}), and typical downstream tasks include sentiment analysis and machine translation. 
In downstream tasks, the FM is frequently called the \textit{backbone} because, although additional learnable layers may be necessary, it holds the bulk of the parameters.

The self-supervised learning (SSL) paradigm is particularly advantageous for HEP because experimental data is unlabelled.
For many tasks in HEP, supervised training is only possible using simulated datasets, where the truth labels are derived from the simulator itself. 
Running high-quality physics simulations~\cite{Agostinelli:2002hh} is a time-consuming process.
Furthermore, these simulations do not perfectly model real data, causing a domain shift between the synthetic samples the model was trained on and the real data to which it is then applied.
Therefore, we are highly motivated to develop SSL techniques for producing FMs that can be trained directly on real data.

In this work, we iterate upon \citet{mpm}, which introduced a SSL strategy designed to run on unordered sets of objects and targeted applications to particles.
The particles are reconstructed objects derived from detector signals captured during a high-energy collision, such as those produced in the Large Hadron Collider (LHC).
The attributes associated with each particle include its kinematics (energy and momentum), particle type, charge, and additional features pertaining to its reconstruction.
In MPM, we are given a set of attributed particles, a random subset is masked, and the model is tasked to reconstruct it.
MPM is analogous to masked language modeling, as in BERT \cite{BERT}, or masked image modeling, as in BEiT \cite{beit}.
But unlike images and text, the particle sets have no natural ordering.

It is possible to frame masked modeling in the context of denoising autoencoders (DAE) \cite{DAE}.
In a DAE, a lossy augmentation is first applied to the inputs, which are then projected via an encoder to a latent space.
A decoder is used to map back to the original uncorrupted signal.
Once the DAE is trained, only the encoder is saved for further applications, while the decoder is typically discarded.
Masking or removing elements from the input sample is a simple, fast, and effective corruption method that underpins many notable models in NLP and CV \cite{BERT, gpt, bart, mae, beit, context_encoders, sdae, data2vec, masked_feat, simmim}.
Masked pre-training requires little prior knowledge of the data and can be applied to a wide variety of fields.
This is the approach taken by MPM.

Many stable particles are produced in any given collision event, which are subsequently captured by the detector. 
However, in this work, we focus on particle jets.
Jets are collimated sprays of particles produced by the hadronization of quarks and gluons.
Multiple jets can be created in an event, and we treat each as a complete set.
The structure and composition of these sets depend highly on the type of particle that produced it.
As an experimental signature of particles with the colored charge, they are key ingredients in studying quantum chromodynamics, the Standard Model, and searches for new physics.
MPM is a method for training an FM, which can either be fine-tuned or used simply as a fixed encoder for various supervised downstream tasks related to the study of jets. 

As most of the particles' features are continuous, we could not naively apply the same successful training strategy as language models like BERT or GPT.
These models predict the full probability distribution function (PDF) for the masked or next token, an embedding that contains rich semantic information~\cite{mae}.
Naive regression methods on continuous variables do not produce the same informative output.
Inspired by the approach used for images in BEiT, the original MPM model, hereto referred to as MPMv1, was trained to recover tokenized representations of each particle derived from a separately trained Vector-Quantized Variational Autoencoder (VQVAE)~\cite{oord2017neural}.
The VQVAE maps the input jet to a set of discreet codebook elements and back again.
Borrowing the language used in BEiT, the VQVAE-encoder is our particle \textit{tokenizer}.
This changes the MPM reconstruction task from regression to classification, as the FM is tasked to predict the codebook ID of the tokenized particle \footnote{  
MPM pre-training could be seen as a knowledge distillation step, where the model has to predict the same latent as the VQVAE, albeit with missing information.}

\citet{mpm} found that using the VQVAE-derived targets during pre-training leads to a more performant FM than direct regression and argued that this was primarily due to two reasons:
(1) The VQVAE latent space is semantically rich, containing high-level abstractions, giving the MPMv1 encoder a more informative target to learn from (this is also the justification used in BEiT).
(2) By changing from a regression to a classification task, the backbone is taught the full conditional posterior distribution rather than just seeking the mean,
which is much more expressive. 
However, producing the VQVAE requires an additional training step in the pipeline.
VQVAEs are notoriously unstable and hard to train.
Furthermore, the aforementioned quantization leads to a loss of information.

In this paper, we make the following contributions. 
(1) We propose an improved MPM training paradigm, named MPMv2, by enhancing model architecture and addressing existing inefficiencies.
We also expand the particle attributes to provide a more detailed representation.
(2) We provide a detailed study of alternative reconstruction tasks for MPMv2 pre-training, ones that replace the costly VQVAE-derived targets.
(3) We provide a new test bed for pre-trained models that include a wider set of downstream tasks commonly encountered in jet physics. 


\section{Related Work}
\label{sec:related}

In addition to MPM, there have been several other works in developing foundation models for physics.
One of the first notable attempts is JetCLR \cite{JetCLR}, which uses the SimCLR \cite{simclr} framework to pre-train a fixed encoder.
JetCLR uses approximate but physically inspired augmentations, such as rotations of the constituents about the jet axis and the smearing of soft constituents to estimate soft gluon radiation.
The SimCLR framework was used again for Re-Simulation-based Self-Supervised Learning (R3SL)~\cite{harris2024resim}.
This framework explicitly requires simulated data as each positive pair is the same underlying event, duplicated at some point in the simulation pipeline, and then completed with different seeds or settings.
OmniJet-$\alpha$ is another recent work that uses a similar approach to MPM but swaps the masked reconstruction pre-training for GPT style next token prediction.
Similar to MPM, \citet{kishimoto2023pretraining} devised a pre-training strategy where only the particle type is masked and reconstructed.
The kinematics and other continuous features are always available to the model. 
The work by \citet{supervisedfinetuning} proposes to describe various elements of the reconstruction pipeline as viable pre-training tasks. 
Finally, the Omnilearn~\cite{mikuni2024omnilearn} model is pre-trained jointly as a supervised classifier for jets and as a diffusion generative model.

\section{Data}
\label{sec:data}

\subsection{Datasets}

A key aspect of MPM is that it does not require labels and can thus be applied directly to experimental data.
However, because large open datasets of real jets are not available, we use MC simulations to refine the framework.
Crucially, we still ignore the truth labels during pre-training, and the only conclusions we draw in this paper are between models trained on the same datasets.
Access to the truth labels also gives us a means to evaluate the performance of the FMs.

We focus on two datasets, both of which utilize the Delphes~\cite{deFavereau:2013fsa} simulation package.
The first is the publicly available JetClass dataset~\cite{JetClass}, which contains 120 million large radius jets equally distributed amongst 10 classes.
Each class represents a different physical process and decay chain, such as $H\rightarrow 4q$ and $t\rightarrow b \ell \nu$.
The second dataset we label BTag, which contains 3 million jets from three classes differentiated by the flavor of quark which initiated the jet, \textit{light, charm}, or \textit{bottom}.

Events in both JetClass and BTag are generated using Pythia8~\cite{Pythia}, but jets in JetClass arising from top, W, Z, or Higgs decays are additionally modeled with MadGraph5~\cite{MadGraph}.
Both datasets reconstruct their jets using calorimeter energy deposits with the anti-kt algorithm~\cite{AntiKt}; the radius parameter is set to $R=0.8$ for JetClass and $R=0.4$ for BTag.
JetClass jets have significantly higher transverse momentum of 500-1000 GeV, whereas BTag only requires $\pt \geq 20$ GeV.
Additionally, JetClass uses a Delphes configuration that matches the CMS experiment~\cite{CMS} while BTag is configured to match the ATLAS experiment~\cite{ATLAS}.
The final significant difference is that JetClass contains both charged and neutral constituents, while BTag only contains charged particles.
As such, JetClass jets have a higher cardinality, averaging around 50 constituents per jet, whereas BTag jets are capped at 15.

We only use JetClass to pre-train our models, but we fine-tune and evaluate using both datasets.
The differences between these datasets represent the realistic variations in how particle physics jets are defined in different experimental settings.
Targeted kinematic ranges, reconstruction parameters (like the anti-kt radius), and object selection vary significantly depending on the physics analyses and are finely tuned by experts. 
These differences offer a chance to view the backbone's generalizability to new downstream tasks and a new out-of-distribution (OOD) dataset.

In \citet{mpm}, each massless constituent is represented using only its kinematics relative to the jet axis, $(\pt, \Delta \eta, \Delta \phi)$.
We expand this to include common reconstructed attributes used in experimental settings.
For charged constituents, which leave tracks in the detector, we include the lifetime signed longitudinal and transverse impact parameters ($d_0$, $z_0$) as well as their reconstruction uncertainties ($\sigma(d_0)$, $\sigma(z_0)$)~\cite{DIPS}.
Neutral particles have no defined impact parameters, so these are zero-padded.
These 7 variables form the continuous features of the particle, \xc.
Also included is the particle identity (ID) \xid, a one-hot encoded vector that categorizes both the particle type and charge into 8 independent classes.
To summarize, a jet is an unordered set of N particles, each represented by a vector of 8 features, 7 continuous and one categorical, $X = \{x_i=(\xc_i, \xid_i)\}_{i=1}^N$.

\section{Method}
\label{sec:eff_mpm}

In MPMv1, $M$ particles out of the $N$ that constitute the jet are selected, and all of their features are replaced with a special masked token.
The goal is then to recover those features, or at least tokenized representations of them.

Framing MPMv1 as a DAE, the input sample $\mathcal{X} = \{x_i\}_{i=1}^N$, its latent projection $\mathcal{Z}$, and the decoder output $\mathcal{D}$ are all sets, so all mappings between them must be permutation equivariant.
Therefore, the encoder is not provided with positional encoding (PE).
Given $\mathcal{X}$, we define the corrupted sample as the union of the surviving subset and a set of identical masked tokens $\mathcal{S}=\{x_i\}_{i=1}^M~\cup~\{m\}_{1}^{N-M}$.
A transformer acts as the encoder, and a multi-layer perceptron (MLP) acts as the decoder, applied separately per set element\footnote{Referred to in \citet{mpm} as the "Masked-Prediction-Head."}.
A consequence of having no PE is that the encoder's outputs corresponding to masked inputs are duplicates.
\citet{mpm} is forced to inject PE based on $\pt$ in the latent space to break this degeneracy for reconstruction while keeping the encoder equivariant.
Each element in $\mathcal{D}$ is then used in the tokenized reconstruction task, where it is compared to the corresponding element of the same jet passed through the encoder of a VQVAE.

We propose a number of alterations to this model for MPMv2.
The repeated use of the same masked token in the encoder means that the transformer layers perform identical operations, wasting computation.
We found that it was significantly more efficient to remove all masked tokens from the input set and reintroduce them only during decoding.
This means that $\mathcal{Z}$ has a lower cardinality than both $\mathcal{X}$ and $\mathcal{D}$.
This change reflects a departure of a model similar to BERT~\cite{BERT} to a model more akin to MAE~\cite{mae}.
As such, we also experimented with expanding the decoder to a full transformer and saw greatly improved results.
The decoder is designed similarly to the encoder, albeit much smaller.
It has one-quarter of the embedding dimension, fewer layers, and fewer attention heads.
With the new decoder, full PE in the latent space provides too much information, trivializing the reconstruction task, which hurts the FM performance.
We find it sufficient to provide PE between the masked elements, not the full jet.
This is achieved by using a unique mask token depending on the \pt~order of the dropped constituents with respect to each other only.
The loss function is then derived by comparing $\mathcal{X}$ and $\mathcal{D}$ in a variety of \textit{reconstruction tasks}.

\begin{figure}[htp!]
    \centering
    \includegraphics[width=0.8\columnwidth]{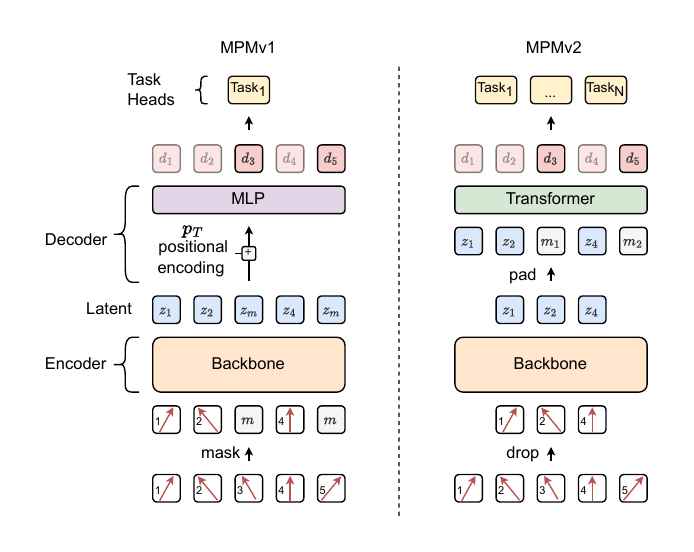}
    \caption{A comparison of the original MPM encoder-decoder setup (left) and the new model configuration (right).
        The new model includes multiple reconstruction tasks, swaps the MLP decoder for a transformer, and only encodes the reduced set.
    }
    \label{fig:mpm1vs2}
\end{figure}

\subsection{Reconstruction Tasks}
\label{sec:recovery}

Where MPMv1 only utilized a VQVAE-derived reconstruction task, we now experiment by combining multiple tasks to recover the continuous and categorical features separately.
Each task requires extra learnable layers (task head) and contributes a loss term, which is summed for the combined pre-training.
We investigate 5 different reconstruction tasks for the continuous features \xc and an extra task for the categorical features \xid.

\subsubsection*{Particle Identification}

The first task is simply to recover the particle type \xid of the dropped constituents.
This is a standard classification problem, so we use a linear layer and the cross-entropy loss function for the task head.

\subsubsection*{VQVAE-Tokenized Classification}
We include the method used in the original MPM work.
A VQVAE is first trained to embed the jet, using only the continuous features, to a set of indices representing the elements in a learned codebook.
We used a codebook size of $1024$ and a codebook vector dimension of 32 following~\citet{vqgan}.
We use a linear layer and the cross-entropy loss function for the task head.


\subsubsection*{Direct Regression}

While \citet{mpm} found direct regression to be insufficient for pre-training, we believe it is worth revisiting owing to the much more powerful decoder.
We use a linear layer and find the best results by using the L1-loss to recover the particles' continuous features. 

\subsubsection*{K-Means Tokenized Classification}

If the VQVAE does not provide a sufficiently semantically rich latent space, its benefit may be simply that it creates a classification task.
Regression is mean-seeking, while the tokenized classification allows us to learn the full conditional posterior of the dropped features, albeit in a discretized form.
To test this, we trial a more trivial token reconstruction task using K-Means centroids.
We fit the K-Means using \xc and the first 1 million jets in JetClass.
Based on preliminary tests, we found that $K=16384$ is the optimal number of centroids.
Fitting the K-Means using the \texttt{torchpq} library \cite{torchpq} took significantly less time than training the VQVAE.
Like the other tasks, we used a single linear layer to map to this space and cross-entropy loss function.


\subsubsection*{Conditional Normalizing Flow}

If the strength of the tokenized form of reconstruction over regression is in learning the full posterior distribution $p(\xc_i|d_i)$, it is possible that we can reproduce this using a generative model.
This also means we do not suffer from the information loss that comes with discretization.
One choice of model is a conditional normalizing flow (CNF) \cite{normflow}, which we implement using the \texttt{normflows} library~\cite{normflows_lib}.
The CNF contains 6 rational-quadratic-spline coupling blocks and a Gaussian base distribution.
Each block contains a two-layer MLP, which outputs the spline parameters for half the features of $\xc_i$ given the other half and the context information $d_i$.
It is trained to maximize the log-likelihood of the transformation.


\subsubsection*{Conditional Flow-Matching}
\label{sec:conddiff}

In recent years, diffusion-based generative models have emerged as the go-to methods for generating high-quality data.
Various frameworks exist that try to generalize and describe this family of models~\cite{song2020,karras2022,flowmatching,variationaldiffusion}.
We follow the conditional flow-matching (CFM) framework from~\citet{flowmatching}.
Here, a model learns the probability vector field between the data distribution and a noise distribution parameterized by time $t \in [0, 1]$.
We consider a time-dependent pdf $p(x, t)$ which connects samples drawn from a data distribution $x_0 \sim p_0(x) = p(x, 0)$ to samples drawn from a noise distribution $\epsilon \sim p_1(x) = p(x, 1)$.
Instead of constructing $p(x, t)$ directly, we could equivalently construct the vector field $u(x, t)$, which relates to the pdf via the continuity equation,
\begin{equation}
    \frac{\partial}{\partial t} p(x, t) = - \nabla \cdot (p(x, t)u(x, t)).
\end{equation}
We use a neural network to approximate the velocity vector $u_\theta \approx u$, where $\theta$ represents the trainable weights.
Directly learning the velocity via the flow-matching objective
\begin{equation}
    L_{FM} = \mathop{\mathbb{E}}_{t, x_t \sim p_t(x)}||u_\theta(x_t, t) - u(x_t, t) ||^2,
\end{equation}
is intractable.
Instead, we can learn the conditional probability paths via the CFM loss,
\begin{equation}
    L_{CFM} = \mathop{\mathbb{E}}_{t, \epsilon \sim p_1, x_t \sim p_t(x|\epsilon)}||u_\theta(x_t, t) - u(x_t, t|\epsilon) ||^2,
\end{equation}
These two objectives are equivalent for network training $\nabla_\theta L_{FM}=\nabla_\theta L_{CFM}$ (under all expectations).
Moreover, $u(x_t, t | \epsilon)$ and $p_t(x|\epsilon)$ do have specific tractable forms. 
One such form is $u(x_t, t | \epsilon) = \frac{\epsilon - x_t}{1-t}$ which leads to Gaussian probability paths.

In practice, we derive the training objective given the continuous features of a particle $\xc_i$ and the corresponding decoder output $d_i$.
We first sample a diffusion time $t$ using the logit-norm distribution from~\citet{stablediffusion3} and sample from the noise distribution $\epsilon \sim \mathcal{N}(0, I)$.
We mix the noise and the original features using a basic linear interpolant to get ${\xc_i}_t = (1-t) \xc_i + t \epsilon$.
The target for the model is the velocity vector $u_i=\xc_i - \epsilon$, which we approximate using a three-layer MLP with a hidden dimension of 256, which takes as inputs ${\xc_i}_t$, $d_i$, and a cosine-embedded form of $t$ following \citet{pcdroid}.
The resulting loss function is written as
\begin{equation}
    L_{CFM} = ||u_\theta( {\xc_i}_t, d_i, t)- (\xc_i-\epsilon)||^2.
\end{equation}


\subsection{Set-to-set Flow-Matching}
\label{sec:sdm}

\begin{figure}[htp!]
    \centering
    \includegraphics[width=0.6\columnwidth]{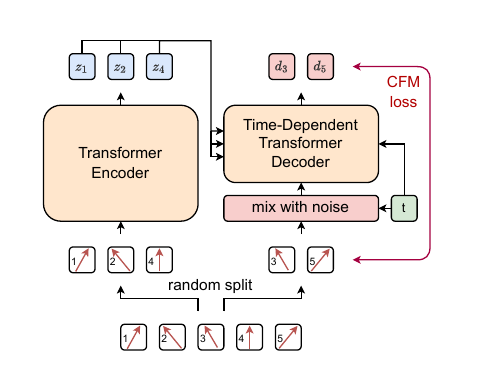}
    \caption{A schematic overview of the SSFM model.}
    \label{fig:mdm}
\end{figure}

We also investigate a set-to-set flow-matching model (SSFM).
The SSFM uses a time-dependent transformer decoder to generate the entire set of constituents given the set of latent nodes.
This setup is similar to the diffusion-masked autoencoder from CV~\cite{diffmae}.
As with MPM, the input set $\mathcal{X}$ is split into a reduced set $\mathcal{S}$ and its complement $\mathcal{T}$.
The reduced set is passed through the encoder to get the latent set $\mathcal{Z}$, which is used in the decoder's cross-attention layers.
The decoder is trained as a set-CFM model to generate the remaining set $\mathcal{T}$.
A diagram of this model is shown in \Cref{fig:mdm}.
Since the loss is based purely on denoising the $\mathcal{T}$, degeneracy is not an issue, and no positional encoding or mask tokens are required.
By varying the masking rate $D_f=\frac{M}{N}$, we can control the amount of jet generated by the diffusion model.
The decoder is a standard diffusion generator when the $D_f=0$.
Thus, our pre-training setup produces a backbone for embedding and a purely generative model for the jets akin to Omnilearn and Omnijet-$alpha$.
During training, we sample $D_f\sim\mathcal{U}(0, 0.8)$ to balance these two objectives.

\section{Results}
\label{sec:experiments}

\subsection{Ablation Studies}
\label{sec:ablation}

To evaluate our proposed alterations to MPMv1, we use the new backbone as a fixed encoder to classify the JetClass dataset.
After pre-training for 200k steps, we freeze the encoder and append a classifier-head, made from 2 class-attention layers \cite{goingdeeper} followed by a linear layer. 
We then train the head as a classifier with cross-entropy loss for another 200k steps.
We elected to use only the regression, K-Means, and particle ID tasks for the ablation study as they were the quickest to prototype.
The full results using all reconstruction tasks are shown in \Cref{sec:downstream}.

We present the results of the ablation study in~\Cref{tab:construction}.
Initially, we recreated the training setup from \citet{mpm}, with the same masking rate of $D_f=0.3$.
Next, we test the setup with more up-to-date transformer layers, described in~\Cref{sec:architecture}.
Then, we add the impact parameters to the features of the particles, followed by including the particle ID inputs and ID reconstruction task.
Each of these steps improves the classification accuracy of both models. 
The largest improvement comes from changing the decoder to a transformer.
This step significantly increased the accuracy of the regression task, bringing the gap between the two methods from 10.5\% to 2.2\%.
To verify the impact of the decoder change, we reran the regression task without the impact parameters or particle ID task.
We found that it achieved an accuracy of 65.0\%, an increase of 9.5\%.
Another major benefit of switching to the MAE setup was a 40\% reduction in GPU memory due to the reduced point cloud size being passed to the encoder.
Finally, we also experimented with adding registers into the encoder \cite{registers}, which prevents the transformer from overwriting elements in the set with global information.
We added 8 registers to the training and found that the classifier's performance increased with little computational cost.
Additionally, we optimized the mask rate and the decoder depth for the final training sessions.

\begin{table}[ht]
    \centering
    \caption{The effects of the model redesign on the accuracy of a classifier head trained using the encoder outputs. All models except the final iteration were trained using 200k training steps, a mask rate of 30\%, and a 2-layer decoder.
    }
    \label{tab:construction}
    \begin{tabular}[t]{lrlrl}
        \toprule
        & \multicolumn{2}{l}{regression} & \multicolumn{2}{l}{k-means}                   \\
        \midrule
        MPMv1 using $(\pt, \eta, \phi)$ \cite{mpm} & 48.9                           &                             & 56.2 &          \\
        + updated transformer layers                                   & 55.5                           & \im{6.6}                    & 62.2 & \im{6.0} \\
        + impact parameter features                           & 62.2                           & \im{6.7}                    & 70.2 & \im{8.0} \\
        + constituent ID feature and ID reconstruction task                   & 63.5                           & \im{1.3}                    & 74.0 & \im{3.8} \\
        + transformer as decoder (MAE)                 & 79.2                           & \im{15.7}                   & 81.4 & \im{7.4} \\
        + registers                                    & 80.4                           & \im{1.2}                    & 83.0 & \im{1.6} \\
        + longer train (1M steps) + deeper decoder + 40\% mask rate                & 83.3                           & \im{2.0}                    & 84.0 & \im{1.0} \\
        \bottomrule
    \end{tabular}
\end{table}


\subsection{Downstream Tasks}
\label{sec:downstream}

Here, we evaluate the performance of our backbones on a variety of downstream tasks typically encountered in jet physics. 
Each backbone is pre-trained using one of the continuous feature reconstruction tasks (which it is named after) together with the particle ID task.
Pre-training is run for 1M steps after which specific downstream task layers are appended to the encoder, and the model is fine-tuned.
Finetuning is run for 200k steps, allowing for early stopping using a validation set.
We use a randomly initialized network as a baseline to highlight the performance provided by pre-training and repeat each experiment 5 times to estimate the run-to-run variance.

\subsubsection{In-Distribution Classification}

We perform classification on the JetClass dataset using the same classifier head described in \Cref{sec:ablation}.
The backbone's data efficiency is measured by varying the number of jets used to train the classifier from 1k to 100M, and these results are shown in \Cref{fig:jetclass}.
At each training set size, the performance of all pre-trained backbones is superior to the randomly initialized network.
However, this boost diminishes as the number of jets increases.
At the maximum 100M jets, all backbones achieve an accuracy between 85.0\% (regression) and 85.3\% (K-Means), whereas the random initialization achieves 84.3\%.
Interestingly, the K-Means backbone performs best with more data, while the CNF and Regression backbones are more data-efficient.
The Flow-backbone achieves the same performance with 10k jets as the randomly initialized network with 1M. 

\subsubsection{Weakly Supervised Classification}

In many experimental settings, we are unable to produce perfectly labeled data, so we are interested in model performance in a setting where the labels are noisy or incomplete.
The principle of \textit{classification without labels} (CWoLa) \cite{cwola} is that the ideal classifier between two mixed datasets with different signal and background proportions is the same as the ideal classifier between the two pure datasets.
This is utilized in template-based anomaly detection \cite{cathode, feta, drapes, notdrapes, curtainsf4f, interplayanomaly} and in muon isolation~\cite{learningisolatemuonsdata}.

\begin{figure}[h!]
    \centering
     \begin{subfigure}{0.32\linewidth}
         \centering
         \includegraphics[width=\linewidth]{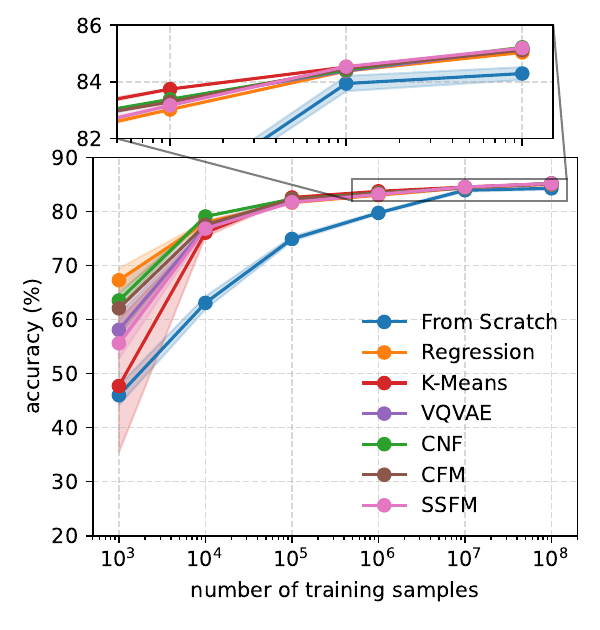}
         \caption{}
         \label{fig:jetclass}
     \end{subfigure}
     \begin{subfigure}[b]{0.32\textwidth}
         \centering
         \includegraphics[width=\linewidth]{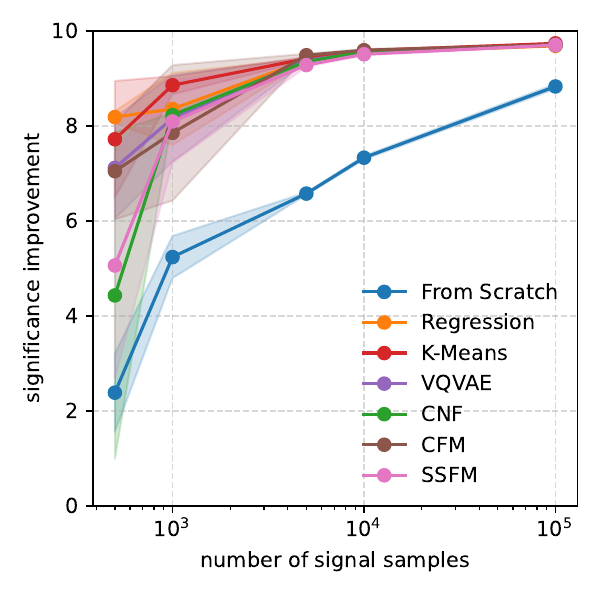}
         \caption{}
         \label{fig:cwola}
     \end{subfigure}
    \caption{The in-distribution performance of the fine-tuned models on the JetClass dataset. (\subref{fig:jetclass}) shows the accuracy using standard supervised classification as a function of the dataset size.  (\subref{fig:cwola}) shows the significance-improvement of the models trained in a CWoLa setting as a function of the number of signal samples in the dataset.}
    \label{fig:plot_A}
\end{figure}

We emulate the CWoLa setting using two datasets of 500k QCD jets.
Into one of the datasets, we inject top-initiated jets as a signal.
We use the same classifier head as in the previous experiments.
In \Cref{fig:cwola}, we show the significance improvement (SIC)~\cite{gallicchio2011} from applying the classifiers on a test set containing pure samples of QCD background and top signal.
The SIC is defined as the signal efficiency (true-positive rate) divided by the square root of the background efficiency (false-positive rate) at a 99\% background rejection.
The pre-trained backbones considerably outperform the benchmark, with the Regression backbone performing the best when only 500 top jets are present in the training set, resulting in a (SIC) of 8.18.

\subsubsection{Out-of-Distribution Classification}

Here, we test the backbones' performance in classifying the BTag dataset, which contains lower-energy, narrower jets with only a few charged particles.
In \Cref{fig:btag}, we show the accuracy of the 3-class classifier as a function of the number of jets used for training.
All pre-trained backbones outperform the benchmark initialization, indicating that the learned mappings are generalizable beyond JetClass.
In this task, the CNF backbone performs the best, but all pre-trained backbones converge at around 70\% accuracy with the maximum number of jets.

\subsubsection{Secondary Vertex Finding}

A track vertex refers to a common point where reconstructed particle tracks originate, indicating the location of an interaction or decay.
Bottom and charm hadrons produced in the collision will survive long enough to travel several millimeters beyond the interaction point before decaying. 
This leads to multiple vertices existing within the same jet, and discovering them is a key intermediate step used in the identification of heavy-flavor jets ~\cite{gn1, svfd}, such as those initiated by bottom and charm hadrons. 
The decay of kaons also causes additional vertices.
Secondary vertex finding is a task that partitions the jet's tracks into groups that all originate from the same vertex. 
It is typically recast as an edge classification task, where given any two tracks, the pair is classified as either being part of the same vertex or not.
This means that for a jet with N tracks, there are $N(N-1)/2$ unique pairs to test. 

The additional layers for this task followed a twin-network approach \cite{siamese}.
Whereby the probability that two tracks $x_i$ and $x_j$ came from the same vertex was defined by $\sigma\left(G\left[|F\left(z_i\right)-F\left(z_j\right)|\right]\right)$, where $G, F$ are MLPs, $z_i, z_j$ are the outputs of encoder, and $\sigma$ is the sigmoid function.
Following~\citet{svfd}, we use the adjusted Rand index (ARI) \cite{ari} as the performance metric.
We plot the ARI as a function of the number of secondary vertices in \Cref{fig:vtx}.
Here, we find that the best-performing model is the backbone trained using the CNF task, though all backbones perform better than the benchmark.

\subsubsection{Heavy Track Identification}

\begin{figure}[h!]
    \centering
     \begin{subfigure}{0.32\linewidth}
         \centering
         \includegraphics[width=\linewidth]{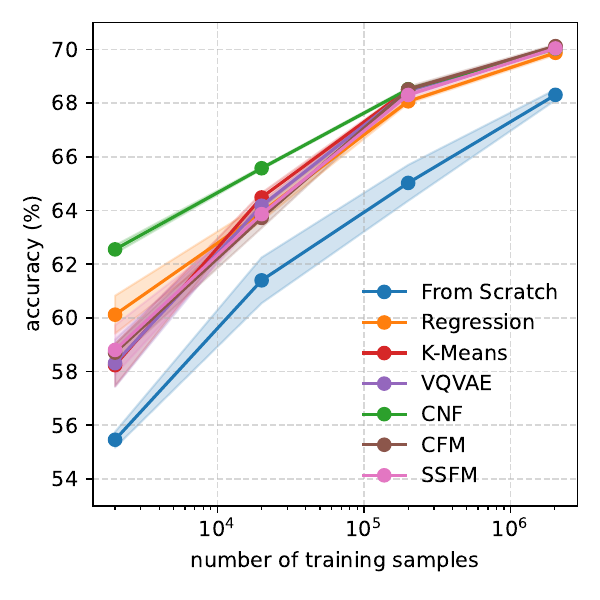}
         \caption{}
         \label{fig:btag}
     \end{subfigure}
     \begin{subfigure}[b]{0.32\textwidth}
         \centering
         \includegraphics[width=\linewidth]{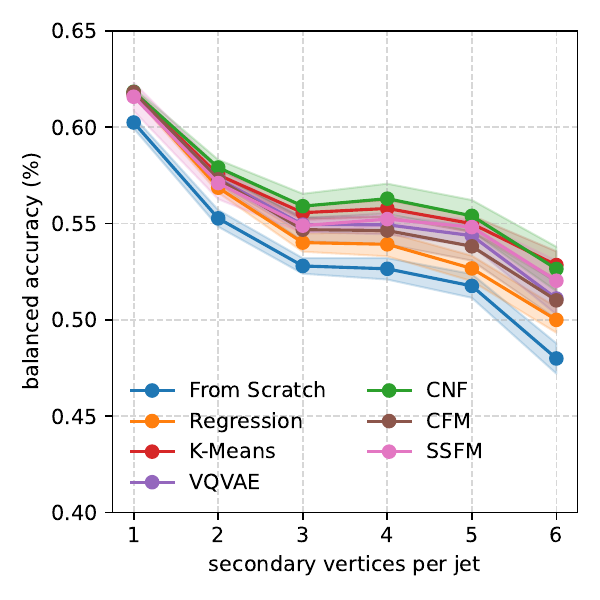}
         \caption{}
         \label{fig:vtx}
     \end{subfigure}
     \begin{subfigure}[b]{0.32\textwidth}
         \centering
         \includegraphics[width=\linewidth]{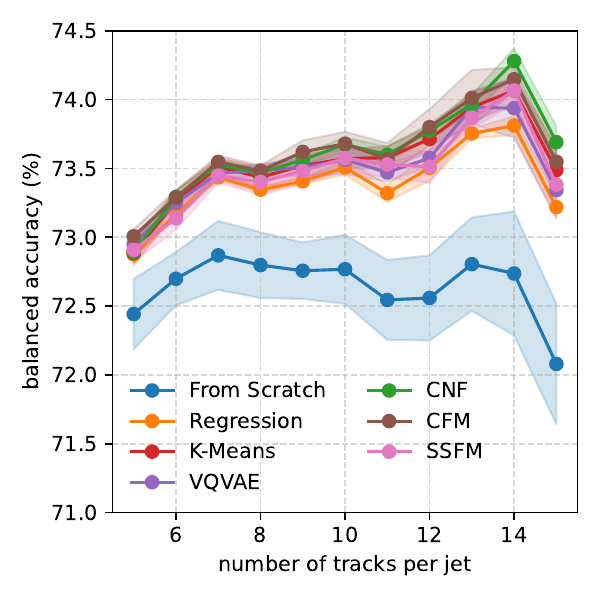}
         \caption{}
         \label{fig:trk}
     \end{subfigure}
    \caption{The performance of the fine-tuned models on the BTag dataset. (\subref{fig:btag}) shows the supervised jet classifier accuracy versus the number of samples used in fine-tuning.
    (\subref{fig:vtx}) shows the ARI score for the segmentation task versus the number of secondary vertices within each jet.  (\subref{fig:trk}) shows the balanced accuracy for the track identification task as a function of the number of tracks in each jet.}
    \label{fig:plot_B}
\end{figure}

Where the vertex finding task grouped tracks that shared a vertex, we can also attempt to identify the type of vertex associated with each track. 
Each of the tracks in the BTag dataset can be associated with having come from a $b$-quark decay, $c$-quark decay, or from the primary vertex (i.e., from heavy quark fragmentation or from light flavor jets).
The head for this task is a simple three-layer MLP attached to the end of the backbone that acts on each of the constituents separately. 
Since the class distributions are so heavily imbalanced, we found that the metric that best highlighted the difference between the pre-training methods was the balanced accuracy.
In \Cref{fig:trk}, we show the balanced accuracy as a function of the number of tracks present in each jet and find that the pre-trained backbones all outperform the baselines.

\section{Conclusion}
\label{sec:conclusion}

In this work, we sought to improve upon the work of \citet{mpm} and answer whether the costly tokenization step is necessary for pre-training.
We achieved this by investigating other methods of reconstruction, including more trivial tokenization via the K-Means algorithm and using conditional generative models.
We have successfully demonstrated that the new models perform considerably better than an untrained backbone and the original MPMv1 in various tasks, including those performed on an OOD dataset. 
We found that the most significant improvement was the adoption of a much more powerful decoder and that the performance between the different continuous reconstruction pre-training tasks was minor. 
We also introduced a new method of pre-training via set-to-set generation, which was highly competitive with MPMv2. 
We believe that these insights demonstrate that we do not require a tokenization step, conclusions which may also affect other SSL models using the VQVAE, such as \citet{birk2024omnijet}.

\section*{Acknowledgements}
TG, SK, and ML, would like to acknowledge funding through the SNSF Sinergia grant CRSII$5\_193716$ called ``Robust Deep Density Models for High-Energy Particle Physics and Solar Flare Analysis (RODEM)'', and the SNSF project grant 200020\_212127 called ``At the two upgrade frontiers: machine learning and the ITk Pixel detector''.
ML also acknowledges the funding acquired through the Swiss Government Excellence Scholarships for Foreign Scholars.
MK is supported by the US Department of Energy (DOE) under grant DE-AC02-76SF00515. LH is supported by the Excellence Cluster ORIGINS, which is funded by the Deutsche Forschungsgemeinschaft (DFG, German Research Foundation) under Germany’s Excellence Strategy - EXC-2094-390783311. MO is supported by USA-Israel BSF - 2022641.

\clearpage

\bibliographystyle{unsrtnat}
\bibliography{main}

\clearpage

\appendix

\section{Model Architecture}
\label{sec:architecture}

We propose a number of alterations to the model introduced by \citet{mpm}, hereto referred to as MPMv1, which was based on the \textit{NormFormer} architecture \cite{normformer}.
We opt for a more standard \textit{pre-norm}~\cite{prenorm} configuration with a transformer encoder comprising 8 layers, each with an embedding dimension of 512.
We use 8 heads for the multi-headed self-attention layers, feedforward network with dimension multipliers of $\times2$, and SwiGLU activations~\cite{glu}.
For both the attention and dense residual updates, we use LayerScale~\cite{goingdeeper}.
The decoder is comprised of the same layer types but is considerably smaller. 
The hyperparameters used are shown in \Cref{tab:hyper}.
All models are trained using the AdamW optimizer with a maximum learning rate of $1 \times 10^{-3}$ and a weight decay of $1 \times 10^{-5}$.
The learning rate schedule was increased linearly from zero over the first 50k steps before exponentially decaying with a half-life of 100k.
All pre-training is performed on the full JetClass training set with a batch size 1000.

\begin{table}[ht]
    \centering
    \caption{Network and training hyperparameters for pre-training the final models.}
    \label{tab:hyper}
    \begin{tabular}{llr}
        \toprule
        \multicolumn{2}{c}{Hyperparameter} & Value \\
        \midrule
        \multirow{5}*{Encoder} & embedding dimension & 512 \\
        & layers & 8 \\
        & attention heads & 8 \\
        & registers & 8 \\
        & activation & SwiGLU \\
        \midrule
        \multirow{5}*{Decoder} & embedding dimension & 128 \\
        & layers & 4 \\
        & attention heads & 4 \\
        & registers & None \\
        & activation & SwiGLU \\
        \midrule
        \multirow{7}*{Training} & optimizer & AdamW \\
        & max learning rate & $1 \times 10^{-3}$ \\
        & weight decay & $1 \times 10^{-5}$ \\
        & batch size & 1000 \\
        & warm-up steps & 50 000 \\
        & training steps & 1 000 000 \\
        & scheduler & exponential \\
        \bottomrule
    \end{tabular}
\end{table}

\section{Data Distributions}
\label{sec:add_plots}

\begin{table}[ht]
\centering
    \caption{The features used to describe each jet constituent.}
    \label{tab:features}
    \begin{tabular}[t]{lr}
        \toprule
        \multicolumn{2}{c}{Continuous features \xc}   \\
        transverse momentum           & $\pt$         \\
        pseudorapidity to jet axis    & $\Delta \eta$ \\
        azimuthal angle to jet axis   & $\Delta \phi$ \\
        transverse impact parameter   & $d_0$         \\
        longitudinal impact parameter & $z_0$         \\
        uncertainty on $d_0$          & $\sigma(d_0)$ \\
        uncertainty on $z_0$          & $\sigma(Z_0)$ \\
        \midrule
        \multicolumn{2}{c}{Particle type \xid}          \\
        photon                        & 0             \\
        negative hadron               & 1             \\
        neutral hadron                & 2             \\
        positive hadron               & 3             \\
        electron                      & 4             \\
        positron                      & 5             \\
        muon                          & 6             \\
        antimuon                      & 7             \\
        \bottomrule
    \end{tabular}
\end{table}

\begin{figure}[h]
    \centering
    \includegraphics[width=0.32\linewidth]{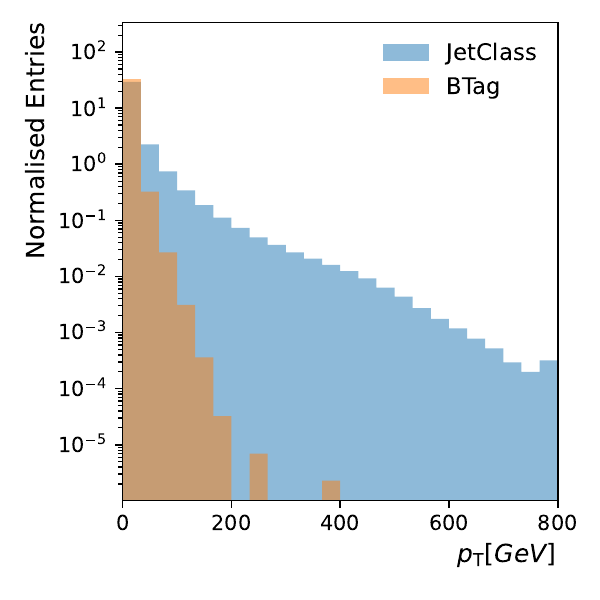}
    \includegraphics[width=0.32\linewidth]{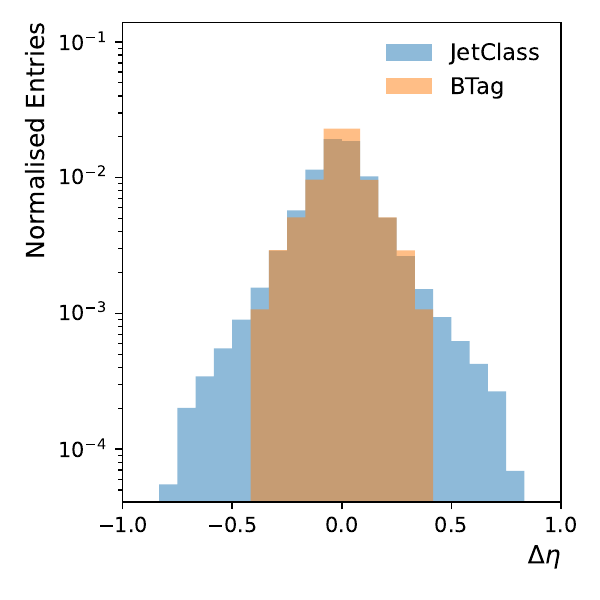}
    \includegraphics[width=0.32\linewidth]{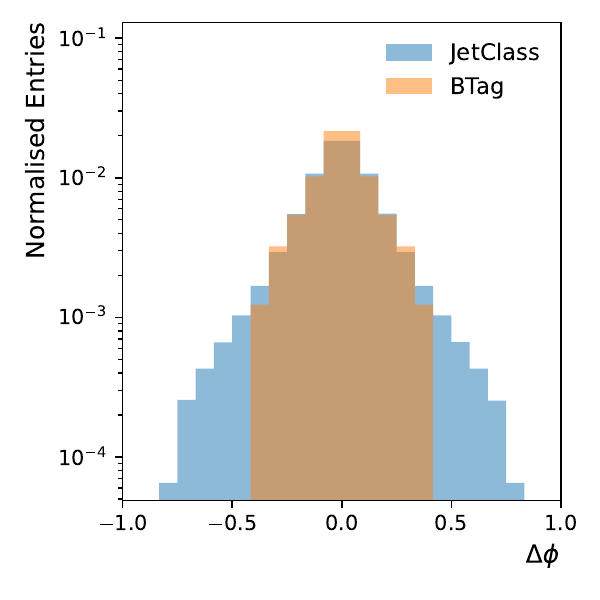}
    \includegraphics[width=0.32\linewidth]{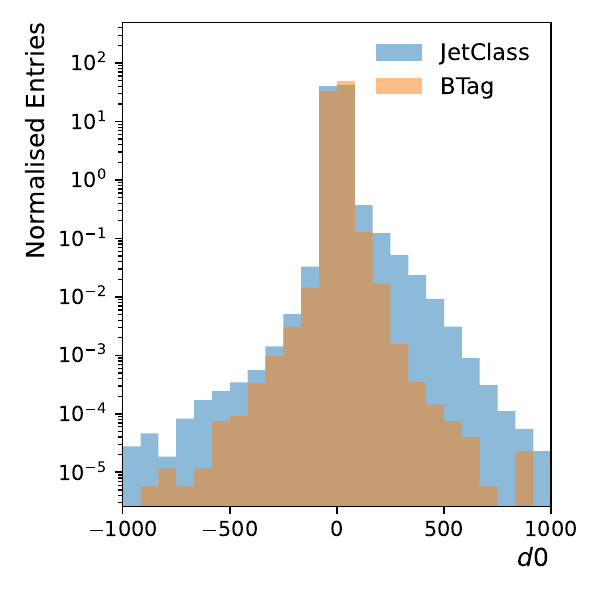}
    \includegraphics[width=0.32\linewidth]{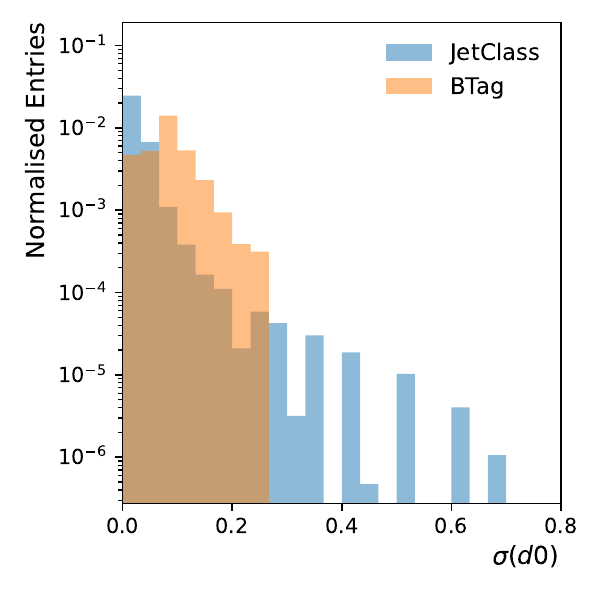}
    \includegraphics[width=0.32\linewidth]{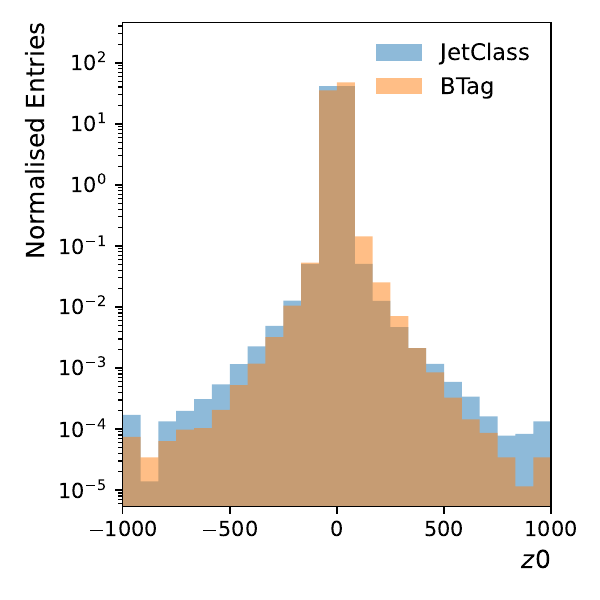}
    \includegraphics[width=0.32\linewidth]{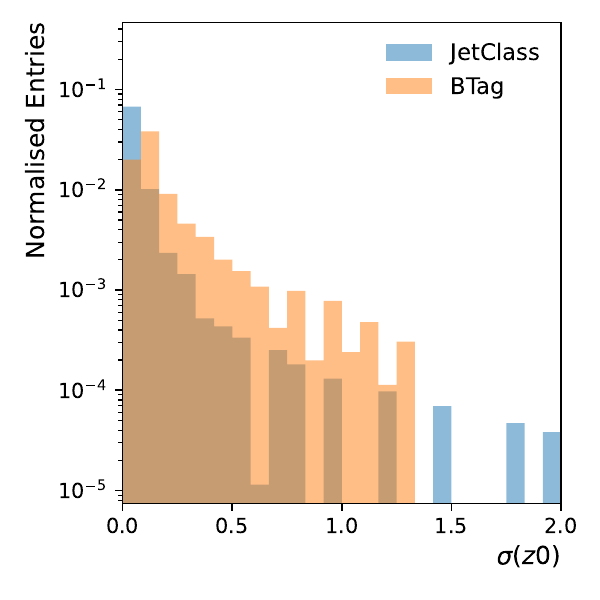}
    \includegraphics[width=0.64\linewidth]{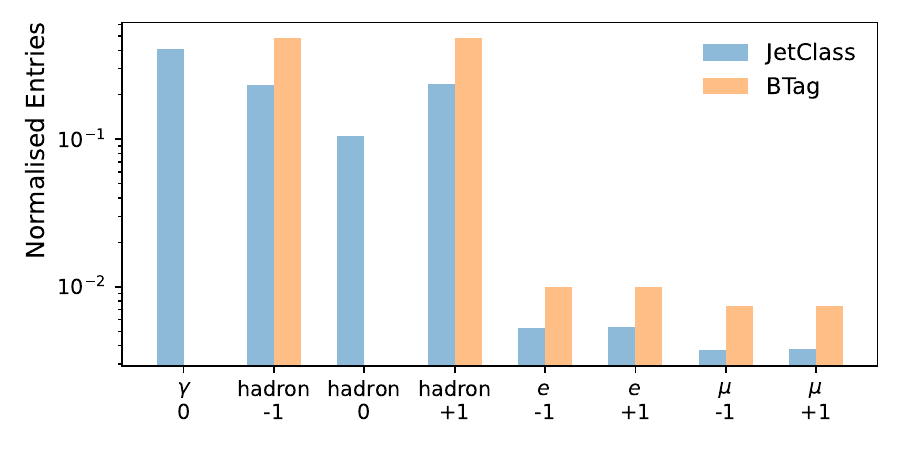}
    \caption{The distributions of the particle features for the two datasets. The final plot shows the distributions of the particle types \xid for the two datasets.}
    \label{fig:continuous}
\end{figure}

\clearpage

\section{Decoder and Mask Rate}

\begin{figure}[htp!]
    \centering
    \includegraphics[width=0.7\linewidth]{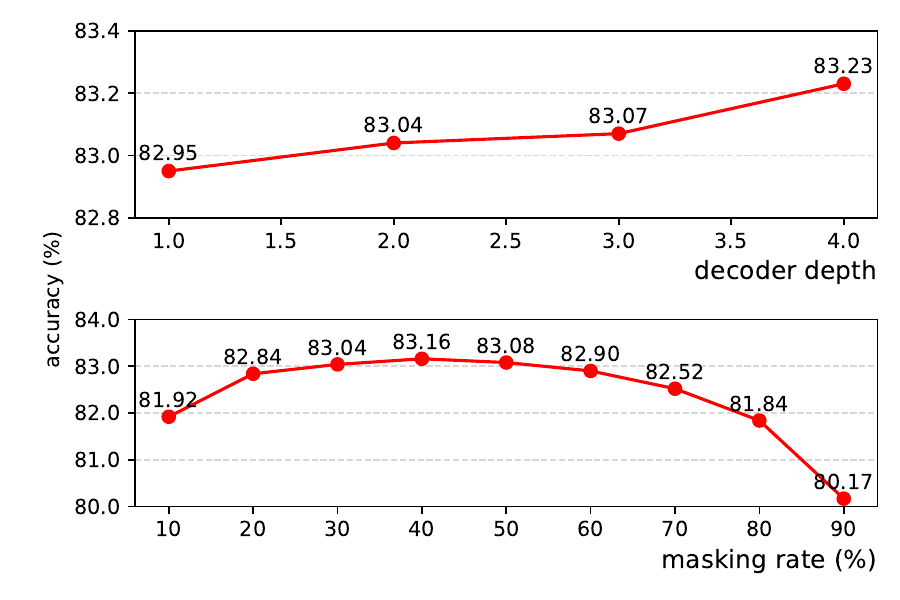}
    \caption{The effect of the decoder depth (top) and the mask rate (bottom) on the classification accuracy using the outputs produced by an MPM backbone trained with the K-Means and ID tasks.}
    \label{fig:sweep}
\end{figure}

Using the K-Means + ID setup, we investigated the effect of the mask rate and the decoder depth.
These results are shown in \Cref{fig:sweep}.
We found that the model was relatively robust to the mask rate but that a rate of 40\% was optimal.
Surprisingly at high levels of masking, 90\%, the model was still able to achieve an accuracy of over 80\%.
We found that increasing the decoder depth improved performance, but due to computational constraints, we explored only up to 4 layers.
We used these optimal settings for the final results.

\section{Fixed Backbone Results}

In addition to fine-tuning, we also investigate the performance of using the frozen pre-trained encoders in the same downstream tasks.
The results are shown in 
This indicates that these backbones indeed provide a feature-rich latent space.

\begin{figure}[h!]
    \centering
     \begin{subfigure}{0.32\linewidth}
         \centering
         \includegraphics[width=\linewidth]{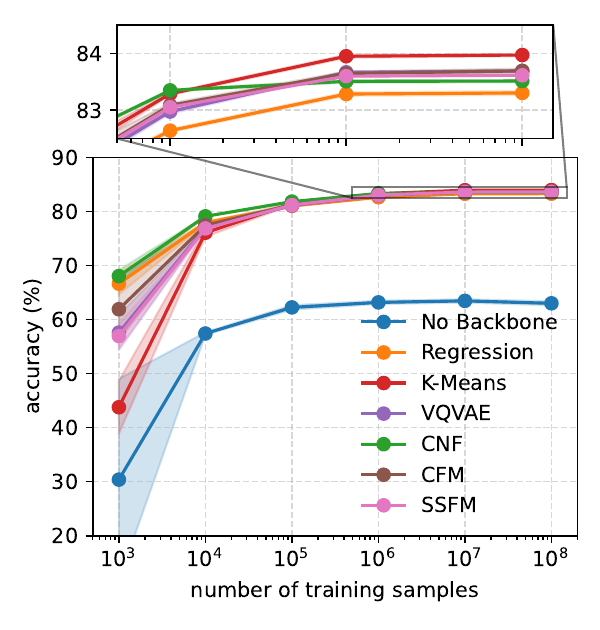}
         \caption{}
         \label{fig:jetclass_fixed}
     \end{subfigure}
     \begin{subfigure}[b]{0.32\textwidth}
         \centering
         \includegraphics[width=\linewidth]{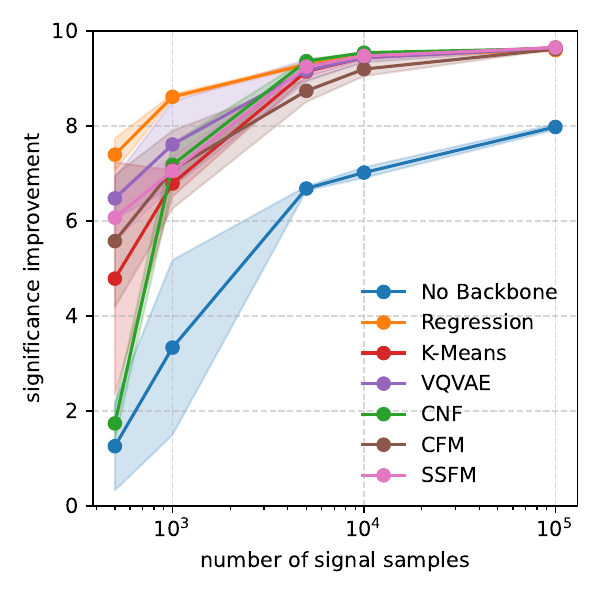}
         \caption{}
         \label{fig:cwola_fixed}
     \end{subfigure}
    \caption{The in-distribution performance of the fixed-backbone models on the JetClass dataset. (\subref{fig:jetclass_fixed}) shows the accuracy using standard supervised classification as a function of the dataset size.  (\subref{fig:cwola_fixed}) shows the significance-improvement of the models trained in a CWoLa setting as a function of the number of signal samples in the dataset.}
    \label{fig:plot_A}
\end{figure}

\begin{figure}[h!]
    \centering
     \begin{subfigure}{0.32\linewidth}
         \centering
         \includegraphics[width=\linewidth]{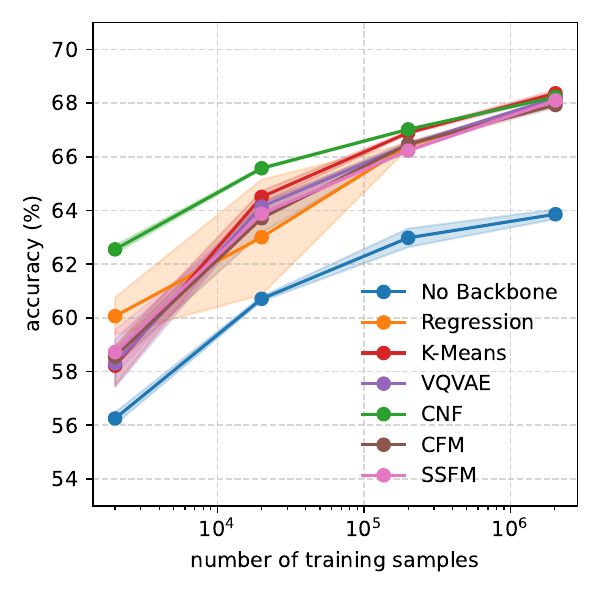}
         \caption{}
         \label{fig:btag_frozen}
     \end{subfigure}
     \begin{subfigure}[b]{0.32\textwidth}
         \centering
         \includegraphics[width=\linewidth]{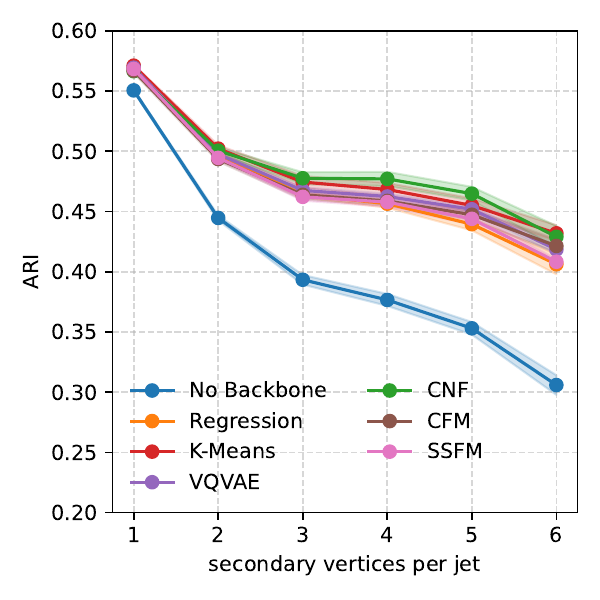}
         \caption{}
         \label{fig:vtx_frozen_ari}
     \end{subfigure}
     \begin{subfigure}[b]{0.32\textwidth}
         \centering
         \includegraphics[width=\linewidth]{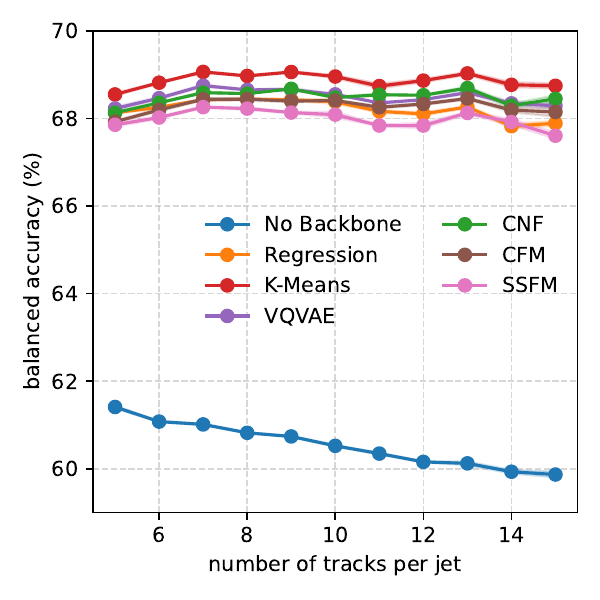}
         \caption{}
         \label{fig:trk_frozen}
     \end{subfigure}
    \caption{The performance of the fixed backbone models on the BTag dataset. (\subref{fig:btag_frozen}) shows the supervised jet classifier accuracy versus the number of samples used in fine-tuning.
    (\subref{fig:vtx_frozen_ari}) shows the ARI score for the segmentation task versus the number of secondary vertices within each jet.  (\subref{fig:trk_frozen}) shows the balanced accuracy for the track identification task as a function of the number of tracks in each jet.}
    \label{fig:plot_B}
\end{figure}

\section{Reconstruction Plots}

\begin{figure}[h]
    \centering
    \includegraphics[width=0.32\linewidth]{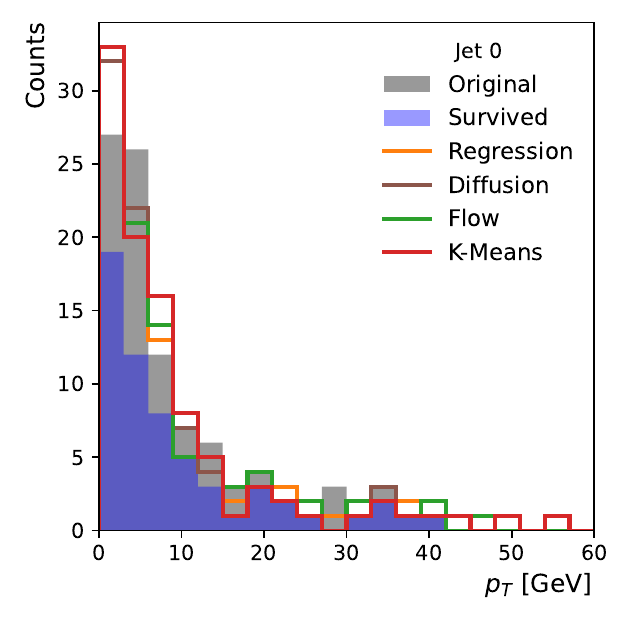}
    \includegraphics[width=0.32\linewidth]{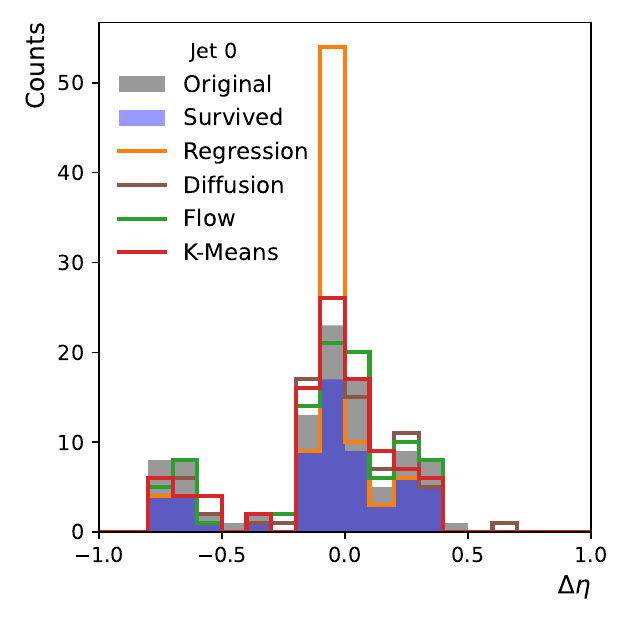}
    \includegraphics[width=0.32\linewidth]{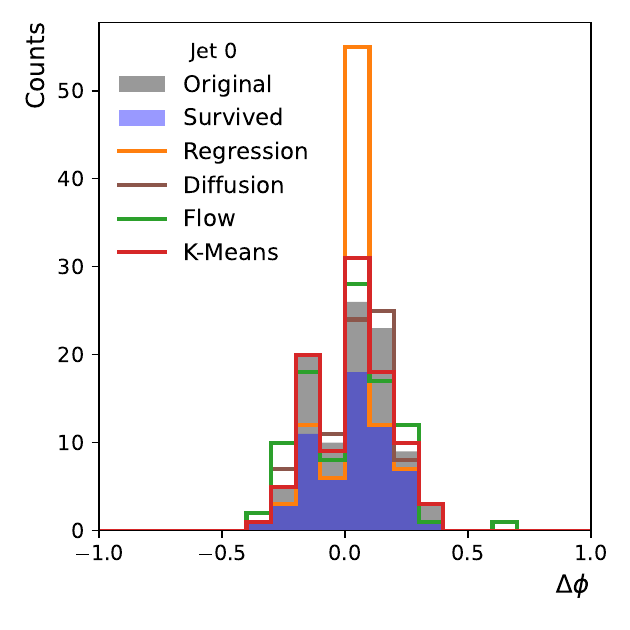}
    \includegraphics[width=0.32\linewidth]{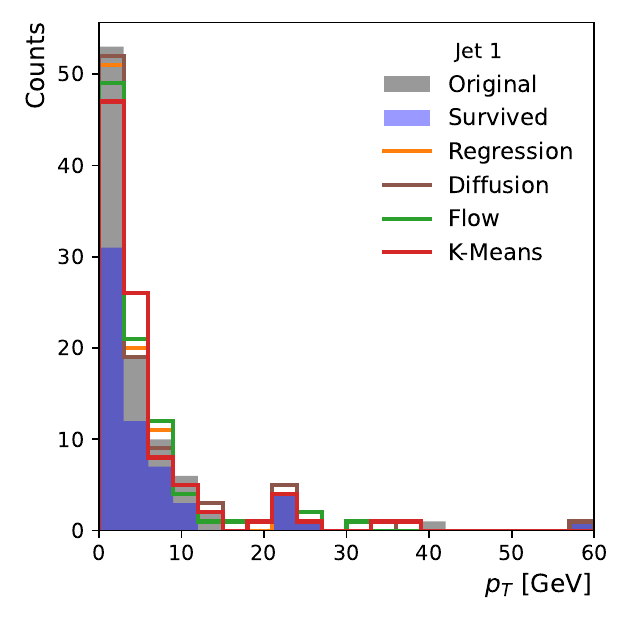}
    \includegraphics[width=0.32\linewidth]{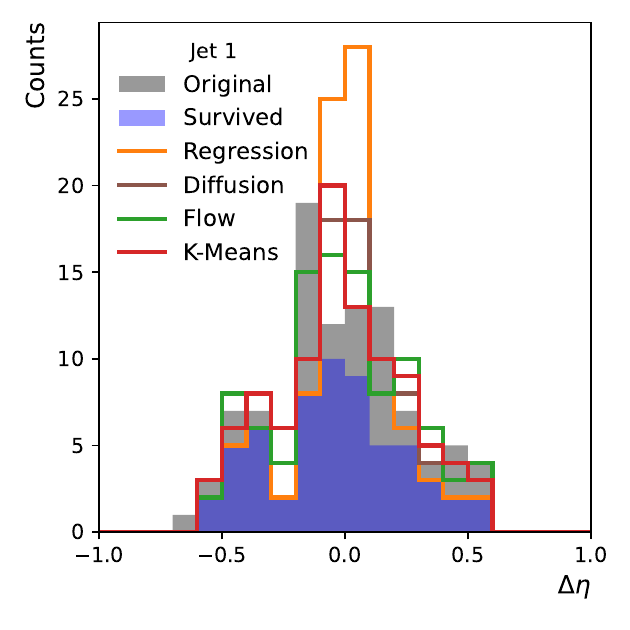}
    \includegraphics[width=0.32\linewidth]{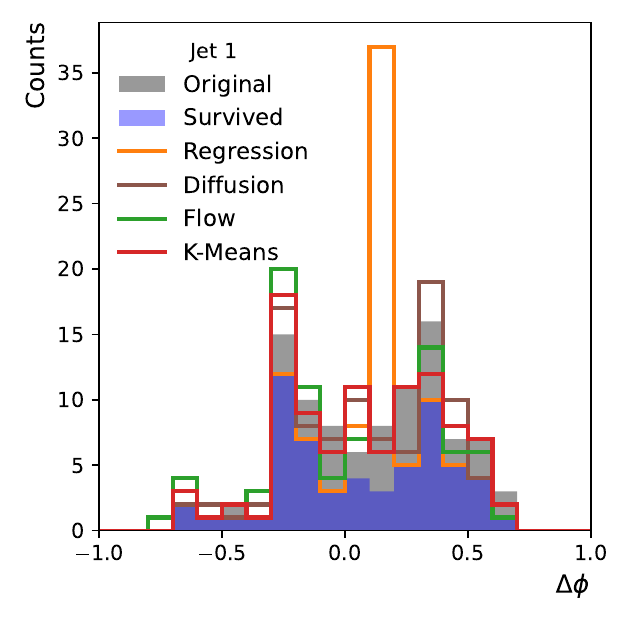}
    \includegraphics[width=0.32\linewidth]{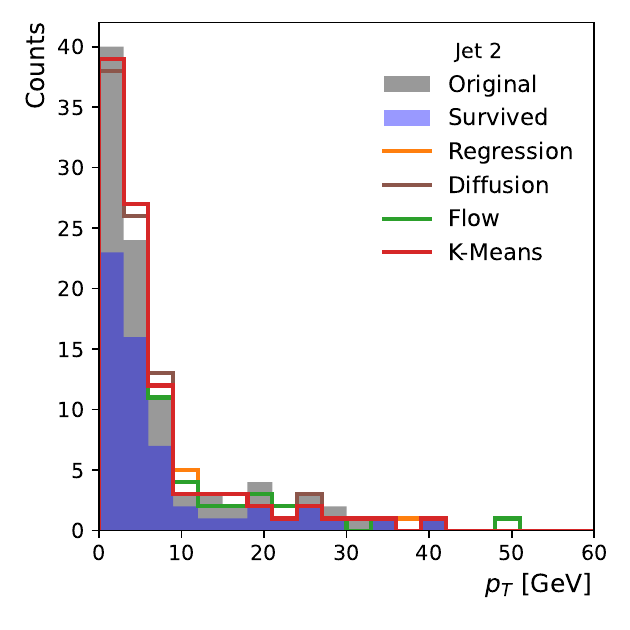}
    \includegraphics[width=0.32\linewidth]{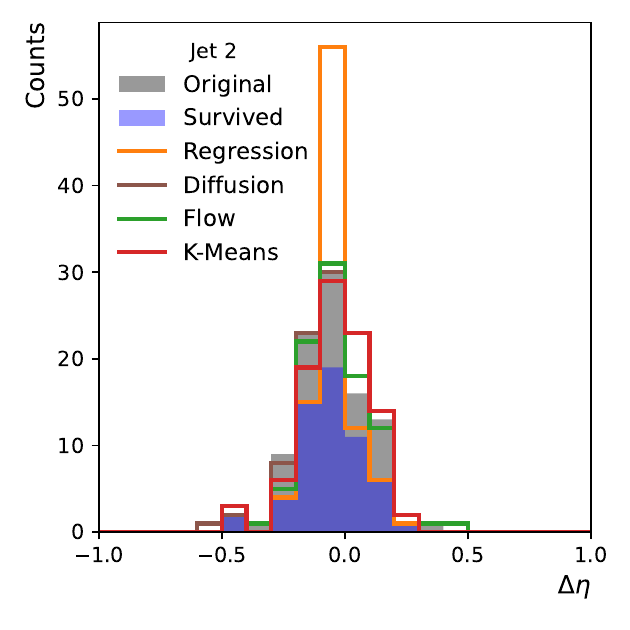}
    \includegraphics[width=0.32\linewidth]{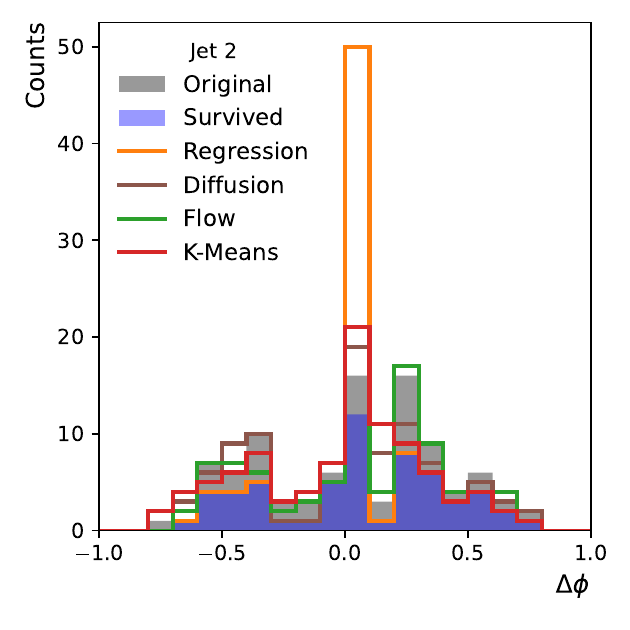}
    \caption{Reconstruction plots for the different backbones. We show 3 randomly selected jets (rows) from the JetClass dataset and plot their $(\pt, \Delta \eta, \Delta \phi)$ distributions (columns). The grey shading shows the original jet distribution, while the blue shading shows the surviving jet distribution after 40\% of the constituents were masked. The colored lines show the reconstructed jets from the different methods. The ideal reconstruction would match the original grey shape.} 
    \label{fig:reconstruction}
\end{figure}

Here we show some qualitative results of some of the continuous reconstruction tasks.
We select 3 jets randomly from the JetClass dataset, perform 40\% masking, and then ask each backbone to reconstruct the dropped constituents. 
For the Regression backbone, we simply take the direct feature predictions. 
For the K-Means backbone, we sample under discrete distribution of centroid probabilities, then take the features of the chosen centroid.
For the CNF backbone, we sample under the normalizing flow.
Finally, for the CFM, we first sample from a Gaussian and then numerically integrate along the predicted trajectories. 
In \Cref{fig:reconstruction}, we see that the Regression backbone often collapses towards the center of the distribution. 
This is most visible for the $\Delta \eta$ distribution of Jet-1, which clearly shows a bi-modal distribution indicative of a dual-prong jet. 
All other methods reconstruct this bi-modality, but the Regression backbone simply predicts the mean. 

\clearpage

%% file: main.bbl
\begin{thebibliography}{66}
\providecommand{\natexlab}[1]{#1}
\providecommand{\url}[1]{\texttt{#1}}
\expandafter\ifx\csname urlstyle\endcsname\relax
  \providecommand{\doi}[1]{doi: #1}\else
  \providecommand{\doi}{doi: \begingroup \urlstyle{rm}\Url}\fi

\bibitem[Golling et~al.(2024{\natexlab{a}})]{mpm}
Tobias Golling et~al.
\newblock Masked particle modeling on sets: Towards self-supervised high energy physics foundation models.
\newblock \emph{Machine Learning: Science and Technology}, 2024{\natexlab{a}}.

\bibitem[Harris et~al.(2024)]{harris2024resim}
Philip Harris et~al.
\newblock Re-simulation-based self-supervised learning for pre-training foundation models, 2024.

\bibitem[Kishimoto et~al.(2023)]{kishimoto2023pretraining}
Tomoe Kishimoto et~al.
\newblock Pre-training strategy using real particle collision data for event classification in collider physics.
\newblock In \emph{Advances in Neural Information Processing Systems}, 2023.

\bibitem[Vigl et~al.(2024)Vigl, Hartman, and Heinrich]{supervisedfinetuning}
Matthias Vigl, Nicole Hartman, and Lukas Heinrich.
\newblock Finetuning foundation models for joint analysis optimization in high energy physics.
\newblock \emph{Machine Learning: Science and Technology}, 5\penalty0 (2):\penalty0 025075, 2024.

\bibitem[Birk et~al.(2024)Birk, Hallin, and Kasieczka]{birk2024omnijet}
Joschka Birk, Anna Hallin, and Gregor Kasieczka.
\newblock Omnijet-$\alpha$: The first cross-task foundation model for particle physics, 2024.

\bibitem[Mikuni and Nachman(2024)]{mikuni2024omnilearn}
Vinicius Mikuni and Benjamin Nachman.
\newblock Omnilearn: A method to simultaneously facilitate all jet physics tasks, 2024.

\bibitem[Zhao et~al.(2024)]{zhao2024large}
Zihan Zhao et~al.
\newblock {Large-Scale Pretraining and Finetuning for Efficient Jet Classification in Particle Physics}.
\newblock In \emph{{22nd International Workshop on Advanced Computing and Analysis Techniques in Physics Research}}, 8 2024.

\bibitem[Dillon et~al.(2022)]{JetCLR}
Barry~M. Dillon et~al.
\newblock {Symmetries, safety, and self-supervision}.
\newblock \emph{SciPost Phys.}, 12\penalty0 (6):\penalty0 188, 2022.

\bibitem[Bommasani et~al.(2022)]{bommasani2022opportunities}
Rishi Bommasani et~al.
\newblock On the opportunities and risks of foundation models, 2022.

\bibitem[Devlin et~al.(2019)]{BERT}
Jacob Devlin et~al.
\newblock Bert: Pre-training of deep bidirectional transformers for language understanding, 2019.

\bibitem[Radford et~al.(2018)]{gpt}
Alec Radford et~al.
\newblock Improving language understanding by generative pre-training, 2018.

\bibitem[OpenAI(2023)]{gpt4}
OpenAI.
\newblock Gpt-4 technical report, 2023.

\bibitem[Lewis et~al.(2019)]{bart}
Mike Lewis et~al.
\newblock Bart: Denoising sequence-to-sequence pre-training for natural language generation, translation, and comprehension, 2019.

\bibitem[Brown et~al.(2020)]{NEURIPS2020_1457c0d6}
Tom Brown et~al.
\newblock Language models are few-shot learners.
\newblock In \emph{Advances in Neural Information Processing Systems}, volume~33, pages 1877--1901, 2020.

\bibitem[Caron et~al.(2021)]{dino}
Mathilde Caron et~al.
\newblock Emerging properties in self-supervised vision transformers.
\newblock In \emph{International Conference on Computer Vision}, pages 9630--9640, 2021.

\bibitem[Ramesh and othersw(2021)]{ramesh2021zeroshot}
Aditya Ramesh and othersw.
\newblock Zero-shot text-to-image generation.
\newblock In \emph{International Conference on Machine Learning}, volume 139, pages 8821--8831, 2021.

\bibitem[Alayrac et~al.(2022)]{AlayracDLMBHLMM22}
Jean{-}Baptiste Alayrac et~al.
\newblock Flamingo: a visual language model for few-shot learning.
\newblock In \emph{Advances in Neural Information Processing Systems}, volume~35, pages 23716--23736, 2022.

\bibitem[Oquab et~al.(2024)]{dinov2}
Maxime Oquab et~al.
\newblock {DINO}v2: Learning robust visual features without supervision.
\newblock \emph{Transactions on Machine Learning Research}, 2024.
\newblock ISSN 2835-8856.

\bibitem[He et~al.(2022)]{mae}
Kaiming He et~al.
\newblock Masked autoencoders are scalable vision learners.
\newblock In \emph{Conference on Computer Vision and Pattern Recognition}, pages 16000--16009, 2022.

\bibitem[Bao et~al.(2022)]{beit}
Hangbo Bao et~al.
\newblock Beit: Bert pre-training of image transformers.
\newblock In \emph{International Conference on Learning Representations}, 2022.

\bibitem[Agostinelli et~al.(2003)]{Agostinelli:2002hh}
S.~Agostinelli et~al.
\newblock {GEANT4: A Simulation toolkit}.
\newblock \emph{Nucl. Instrum. Meth. A.}, A506:\penalty0 250--303, 2003.
\newblock \doi{10.1016/S0168-9002(03)01368-8}.

\bibitem[Vincent et~al.(2008)]{DAE}
Pascal Vincent et~al.
\newblock Extracting and composing robust features with denoising autoencoders.
\newblock In \emph{International Conference on Machine Learning}, pages 1096--1103, 2008.

\bibitem[Pathak et~al.(2016)]{context_encoders}
Deepak Pathak et~al.
\newblock Context encoders: Feature learning by inpainting.
\newblock In \emph{Conference on Computer Vision and Pattern Recognition}, pages 2536--2544, 2016.

\bibitem[Vincent et~al.(2010)]{sdae}
Pascal Vincent et~al.
\newblock Stacked denoising autoencoders: Learning useful representations in a deep network with a local denoising criterion.
\newblock \emph{Journal of Machine Learning Research}, 11\penalty0 (12), 2010.
\newblock URL \url{http://jmlr.org/papers/v11/vincent10a.html}.

\bibitem[Baevski et~al.(2022)]{data2vec}
Alexei Baevski et~al.
\newblock Data2vec: A general framework for self-supervised learning in speech, vision and language.
\newblock In \emph{International Conference on Machine Learning}, pages 1298--1312, 2022.

\bibitem[Wei et~al.(2022)]{masked_feat}
Chen Wei et~al.
\newblock Masked feature prediction for self-supervised visual pre-training.
\newblock In \emph{Conference on Computer Vision and Pattern Recognition}, pages 14668--14678, 2022.

\bibitem[Xie et~al.(2022)]{simmim}
Zhenda Xie et~al.
\newblock Simmim: A simple framework for masked image modeling.
\newblock In \emph{Conference on Computer Vision and Pattern Recognition}, pages 9653--9663, 2022.

\bibitem[van~den Oord et~al.(2017)van~den Oord, Vinyals, and Kavukcuoglu]{oord2017neural}
Aaron van~den Oord, Oriol Vinyals, and Koray Kavukcuoglu.
\newblock Neural discrete representation learning.
\newblock In \emph{Advances in Neural Information Processing Systems}, volume~30, 2017.

\bibitem[Chen et~al.(2020)]{simclr}
Ting Chen et~al.
\newblock A simple framework for contrastive learning of visual representations.
\newblock In \emph{International Conference on Machine Learning}, pages 1597--1607, 2020.

\bibitem[de~Favereau et~al.(2014)]{deFavereau:2013fsa}
J.~de~Favereau et~al.
\newblock {DELPHES 3, A modular framework for fast simulation of a generic collider experiment}.
\newblock \emph{JHEP}, 02:\penalty0 057, 2014.

\bibitem[Qu et~al.(2022)Qu, Li, and Qian]{JetClass}
Huilin Qu, Congqiao Li, and Sitian Qian.
\newblock {JetClass: A Large-Scale Dataset for Deep Learning in Jet Physics}, 2022.
\newblock URL \url{https://doi.org/10.5281/zenodo.6619768}.

\bibitem[Sjöstrand et~al.(2008)Sjöstrand, Mrenna, and Skands]{Pythia}
Torbjörn Sjöstrand, Stephen Mrenna, and Peter Skands.
\newblock A brief introduction to pythia 8.1.
\newblock \emph{Comput. Phys. Commun.}, 178:\penalty0 852--867, 2008.

\bibitem[Alwall et~al.(2014)]{MadGraph}
Johan Alwall et~al.
\newblock The automated computation of tree-level and next-to-leading order differential cross sections, and their matching to parton shower simulations.
\newblock \emph{JHEP}, 07:\penalty0 79, 2014.

\bibitem[Cacciari et~al.(2008)Cacciari, Salam, and Soyez]{AntiKt}
Matteo Cacciari, Gavin~P Salam, and Gregory Soyez.
\newblock The anti-kt jet clustering algorithm.
\newblock \emph{JHEP}, 04:\penalty0 063, 2008.

\bibitem[{CMS Collaboration}(2008)]{CMS}
{CMS Collaboration}.
\newblock {The CMS experiment at the CERN LHC}.
\newblock \emph{Journal of Instrumentation}, 3\penalty0 (08):\penalty0 S08004, 2008.
\newblock \doi{10.1088/1748-0221/3/08/S08004}.

\bibitem[{ATLAS Collaboration}(2008)]{ATLAS}
{ATLAS Collaboration}.
\newblock {The ATLAS Experiment at the CERN Large Hadron Collider}.
\newblock \emph{Journal of Instrumentation}, 3\penalty0 (08):\penalty0 S08003, 2008.
\newblock \doi{10.1088/1748-0221/3/08/S08003}.
\newblock URL \url{https://dx.doi.org/10.1088/1748-0221/3/08/S08003}.

\bibitem[{ATLAS Collaboration}(2020)]{DIPS}
{ATLAS Collaboration}.
\newblock {Deep Sets based Neural Networks for Impact Parameter Flavour Tagging in ATLAS}.
\newblock tech. report, {CERN}, 2020.
\newblock URL \url{https://cds.cern.ch/record/2718948}.

\bibitem[Yu et~al.(2022)]{vqgan}
Jiahui Yu et~al.
\newblock Vector-quantized image modeling with improved {VQGAN}.
\newblock In \emph{International Conference on Learning Representations}, 2022.

\bibitem[Omer(2021)]{torchpq}
Sehban Omer.
\newblock {TorchPQ}, 2021.
\newblock URL \url{https://github.com/DeMoriarty/TorchPQ}.

\bibitem[Rezende and Mohamed(2015)]{normflow}
Danilo Rezende and Shakir Mohamed.
\newblock Variational inference with normalizing flows.
\newblock In \emph{International Conference on Machine Learning}, pages 1530--1538, 2015.

\bibitem[Stimper et~al.(2023)]{normflows_lib}
Vincent Stimper et~al.
\newblock normflows: A pytorch package for normalizing flows.
\newblock \emph{Journal of Open Source Software}, 8\penalty0 (86):\penalty0 5361, 2023.

\bibitem[Song et~al.(2020)]{song2020}
Yang Song et~al.
\newblock Score-based generative modeling through stochastic differential equations, 2020.

\bibitem[Karras et~al.(2022)]{karras2022}
Tero Karras et~al.
\newblock Elucidating the design space of diffusion-based generative models, 2022.

\bibitem[Lipman et~al.(2023)]{flowmatching}
Yaron Lipman et~al.
\newblock Flow matching for generative modeling.
\newblock In \emph{International Conference on Learning Representations}, 2023.

\bibitem[Kingma et~al.(2023)]{variationaldiffusion}
Diederik~P. Kingma et~al.
\newblock Variational diffusion models, 2023.

\bibitem[Sauer et~al.(2024)]{stablediffusion3}
Axel Sauer et~al.
\newblock Fast high-resolution image synthesis with latent adversarial diffusion distillation, 2024.

\bibitem[Leigh et~al.(2024)]{pcdroid}
Matthew Leigh et~al.
\newblock Faster diffusion model with improved quality for particle cloud generation.
\newblock \emph{Phys. Rev. D}, 109:\penalty0 012010, 2024.

\bibitem[Wei et~al.(2023)]{diffmae}
Chen Wei et~al.
\newblock Diffusion models as masked autoencoders.
\newblock In \emph{International Conference on Computer Vision}, pages 16284--16294, 2023.

\bibitem[Touvron et~al.(2021)]{goingdeeper}
Hugo Touvron et~al.
\newblock Going deeper with image transformers.
\newblock In \emph{International Conference on Computer Vision}, pages 32--42, 2021.

\bibitem[Darcet et~al.(2024)]{registers}
Timothée Darcet et~al.
\newblock Vision transformers need registers.
\newblock In \emph{International Conference on Learning Representations}, 2024.

\bibitem[Metodiev et~al.(2017)Metodiev, Nachman, and Thaler]{cwola}
Eric~M Metodiev, Benjamin Nachman, and Jesse Thaler.
\newblock Classification without labels: Learning from mixed samples in high energy physics.
\newblock \emph{Journal of High Energy Physics}, 2017\penalty0 (10):\penalty0 1--18, 2017.

\bibitem[Hallin et~al.(2022)]{cathode}
Anna Hallin et~al.
\newblock Classifying anomalies through outer density estimation (cathode).
\newblock \emph{Phys. Rev. D}, 106:\penalty0 055006, 2022.

\bibitem[Golling(2023)]{feta}
Tobias~andothers Golling.
\newblock {Flow-enhanced transportation for anomaly detection}.
\newblock \emph{Phys. Rev. D}, 107\penalty0 (9):\penalty0 096025, 2023.

\bibitem[Sengupta et~al.(2024{\natexlab{a}})]{drapes}
Debajyoti Sengupta et~al.
\newblock {Improving new physics searches with diffusion models for event observables and jet constituents}.
\newblock \emph{JHEP}, 04:\penalty0 109, 2024{\natexlab{a}}.
\newblock \doi{10.1007/JHEP04(2024)109}.

\bibitem[Buhmann et~al.(2024)]{notdrapes}
Erik Buhmann et~al.
\newblock {Full phase space resonant anomaly detection}.
\newblock \emph{Phys. Rev. D}, 109\penalty0 (5):\penalty0 055015, 2024.

\bibitem[Sengupta et~al.(2024{\natexlab{b}})]{curtainsf4f}
Debajyoti Sengupta et~al.
\newblock {CURTAINs flows for flows: Constructing unobserved regions with maximum likelihood estimation}.
\newblock \emph{SciPost Phys.}, 17:\penalty0 046, 2024{\natexlab{b}}.

\bibitem[Golling et~al.(2024{\natexlab{b}})]{interplayanomaly}
Tobias Golling et~al.
\newblock {The Interplay of Machine Learning--based Resonant Anomaly Detection Methods}.
\newblock \emph{Eur. Phys. J. C.}, 84, 03 2024{\natexlab{b}}.

\bibitem[Witkowski et~al.(2023)Witkowski, Nachman, and Whiteson]{learningisolatemuonsdata}
Edmund Witkowski, Benjamin Nachman, and Daniel Whiteson.
\newblock Learning to isolate muons in data.
\newblock \emph{Phys. Rev. D}, 108:\penalty0 092008, 2023.

\bibitem[Gallicchio et~al.(2011)]{gallicchio2011}
Jason Gallicchio et~al.
\newblock Multivariate discrimination and the higgs+w/z search.
\newblock \emph{Journal of High Energy Physics}, \penalty0 (4):\penalty0 69, 2011.

\bibitem[{ATLAS Collaboration}(2022)]{gn1}
{ATLAS Collaboration}.
\newblock {Graph Neural Network Jet Flavour Tagging with the ATLAS Detector}.
\newblock tech. report, {CERN}, 2022.
\newblock URL \url{https://cds.cern.ch/record/2811135}.

\bibitem[Shlomi et~al.(2021)]{svfd}
Jonathan Shlomi et~al.
\newblock {Secondary vertex finding in jets with neural networks}.
\newblock \emph{Eur. Phys. J. C}, 81\penalty0 (6):\penalty0 540, 2021.

\bibitem[Koch et~al.(2015)]{siamese}
Gregory Koch et~al.
\newblock Siamese neural networks for one-shot image recognition.
\newblock In \emph{International Conference on Machine Learning}, volume~2, pages 1--30, 2015.

\bibitem[Hubert and Arabie(1985)]{ari}
Lawrence Hubert and Phipps Arabie.
\newblock {Comparing partitions}.
\newblock \emph{Journal of Classification}, 2\penalty0 (1):\penalty0 193--218, 1985.
\newblock \doi{10.1007/BF01908075}.

\bibitem[Shleifer et~al.(2021)]{normformer}
Sam Shleifer et~al.
\newblock Normformer: Improved transformer pretraining with extra normalization, 2021.

\bibitem[Xiong et~al.(2020)]{prenorm}
Ruibin Xiong et~al.
\newblock On layer normalization in the transformer architecture.
\newblock In \emph{International Conference on Machine Learning}, pages 10524--10533, 2020.

\bibitem[Shazeer(2020)]{glu}
Noam Shazeer.
\newblock Glu variants improve transformer, 2020.

\end{thebibliography}
